\documentclass[aip,cha,amsmath,amssymb,reprint]{revtex4-1}

\usepackage[utf8]{inputenc}
\usepackage[T1]{fontenc}
\usepackage{mathptmx}
\usepackage{etoolbox}

\usepackage{hyperref}
\usepackage{graphicx}
\usepackage{dcolumn}

\makeatletter
\def\@email#1#2{%
 \endgroup
 \patchcmd{\titleblock@produce}
  {\frontmatter@RRAPformat}
  {\frontmatter@RRAPformat{\produce@RRAP{#1\href{mailto:#2}{#2}}}\frontmatter@RRAPformat}
  {}{}
}%
\makeatother

\usepackage{xcolor}

\newcommand{\im}[0]{{i}}
\newcommand{\modd}[1]{\left| #1\right|}
\newcommand{\mb}[1]{\mathbf{#1}}
\newcommand{\mc}[1]{\mathcal{#1}}

\newcommand{\sinc}[0]{\text{sinc}}
\newcommand{\rect}[0]{\text{rect}}
\newcommand{\floor}[1]{\left\lfloor {#1} \right\rfloor}
\begin{document}

\title{Unwrapping photonic reservoirs: enhanced expressivity via random Fourier encoding over stretched domains}

\author{Gerard McCaul}
\affiliation{Department of Physics, Loughborough University, LE11 3TU, Loughborough, United Kingdom}
\author{Girish Tripathy}
\affiliation{Emergent Photonics Research Centre, Department of Physics, Loughborough University, LE11 3TU, Loughborough, United Kingdom}
\author{Giulia Marcucci$^*$}
\affiliation{{LumiAIres Ltd}, {{Glasgow}, {G12 0DG}, {United Kingdom}}}
\author{Juan Sebastian Totero Gongora $^\dagger$}
\email[Corresponding Authors:]{$^\dagger$j.totero-gongora@lboro.ac.uk; $^*$giulia@lumiaires.co.uk}
\affiliation{Emergent Photonics Research Centre, Department of Physics, Loughborough University, LE11 3TU, Loughborough, United Kingdom}
\date{\today}

\begin{abstract}
Photonic Reservoir Computing (RC) systems leverage the complex propagation and nonlinear interaction of optical waves to perform information processing tasks. These systems employ a combination of optical data encoding (in the field amplitude and/or phase), random scattering, and nonlinear detection to generate nonlinear features that can be processed via a linear readout layer. 
In this work, we propose a novel scattering-assisted photonic reservoir encoding scheme where the input phase is deliberately wrapped multiple times beyond the natural period of the optical waves $[0,2\pi)$.
We demonstrate that, rather than hindering nonlinear separability through loss of bijectivity, wrapping significantly improves the reservoir's prediction performance across regression and classification tasks that are unattainable within the canonical $2\pi$ period. 
We demonstrate that this counterintuitive effect stems from the nonlinear interference between sets of random synthetic frequencies introduced by the encoding, which generates a rich feature space spanning both the feature and sample dimensions of the data. 
Our results highlight the potential of engineered phase wrapping as a computational resource in RC systems based on phase encoding, paving the way for novel approaches to designing and optimizing physical computing platforms based on topological and geometric stretching. 
\end{abstract}

\pacs{XXXXX}

\maketitle 

\begin{quotation}
    Physical reservoir computing (RC) systems leverage the dynamics of nonlinear physical systems to perform information processing tasks. 
    In Photonics, a quintessential demonstration of the unique scalability advantages enabled by physical processing is found in systems combining wavefront shaping, random wave propagation, and field intensity detection. In these systems, the input data is encoded into the phase of an optical field, which then propagates through a scattering medium and is detected through a camera. 
    Traditionally, phase encoding has been restricted to the natural period of optical waves, $[0, 2\pi)$, to ensure the separability of outputs. In this work, we relax this constraint and investigate the impact of \textit{phase wrapping} on the performance of a scattering-assisted, feed-forward photonic reservoir. By extending the phase encoding domain beyond its natural period, we demonstrate that the reservoir's prediction performance is significantly improved across regression and classification tasks that are unattainable within a single period domain. We demonstrate that this effect originates from the nonlinear interference between sets of random synthetic frequencies introduced by the encoding, which generate a rich feature space for accurate linear readout. Our results highlight the potential of phase wrapping as a computational resource in photonic RC systems, paving the way for novel approaches to designing and optimizing systems that leverage topological and geometric expansion of feature space through the manipulation of optical waves.  
\end{quotation}

\section{Introduction}
Reservoir Computing (RC) and its feed-forward counterpart, Extreme Learning Machine (ELM), present an efficient framework for harnessing the dynamics of complex nonlinear systems to perform information processing tasks \cite{Gauthier2021, jaegerHarnessingNonlinearityPredicting2004a, maassRealTimeComputingStable2002, daleRoleStructureComplexity2019}. By encoding a signal at the input of a high-dimensional dynamical system — the ``reservoir" —  information is non-linearly mapped into a feature space. This dimensional extension of data then allows it to be read out from a final linear regression layer. In this sense, RC circumvents the computationally intensive backpropagation common in traditional deep neural networks \cite{marcucciTheoryNeuromorphicComputing2020, Dong2018ScalingRC, LukoseviciusPracticalGuideApplying2012}, and instead offloads the computational burden onto physical dynamics.  

Lifting the requirement for backpropagation \textit{in materio}, which would require the availability of scalable and tuneable hardware controllers, RC is particularly attractive for the development of analog and neuromorphic processors \cite{tanakaRecentAdvancesPhysical2019,kudithipudiNeuromorphicComputingScale2025, markovicPhysicsNeuromorphicComputing2020,roySpikebasedMachineIntelligence2019,xuLargescalePhotonicChiplet2024}. 
Several platforms have been explored as realizations of reservoir computers, in both the classical and quantum regimes \cite{Fujii2017, Ghosh2019, polaritoniccomputing, nokkala_gaussian_2021, nokkala_online_2023, Govia2021QuantumPhaseRC, angelatos_reservoir_2021, mccaul_towards_2023, Kalfus2021, mccaulMinimalQuantumReservoirs2025,gartsideReconfigurableTrainingReservoir2022,leeTaskadaptivePhysicalReservoir2024, nerenbergPhotonNumberresolvingQuantum2025}. A particularly promising dynamical substrate is provided by nonlinear wave propagation \cite{marcucciTheoryNeuromorphicComputing2020,hughesWavePhysicsAnalog2019a}. This is reflected in the rapid expension of research targeting \textit{optical} reservoir computing \cite{wangUltrafastSiliconPhotonic2024,brunnerRoadmapNeuromorphicPhotonics2025,mcmahonPhysicsOpticalComputing2023,pierangeliPhotonicExtremeLearning2021,jauriguePostprocessingMethodsDelay2025a}. This approach offers compelling advantages over conventional digital computation, including inherent parallelism, high processing speeds, and significantly lower power dissipation \cite{ShastriPhotonicsArtificialIntelligence2021,marcucciTheoryNeuromorphicComputing2020,valensiseLargescalePhotonicNatural2022,wangOpticalTrainingLargescale2024,wangOpticalNextGeneration2024,olivieriAdiabaticEnergeticAnnealing2025,talukderSpikingPhotonicNeural2024}. To achieve the complexity necessary to act as reservoirs, these systems leverage complex light propagation and interactions, such as multiple scattering, phase modulation and nonlinear frequency mixing, to deliver the necessary complex, high-dimensional transformations required for RC operation. 

In this context, a quintessential example of the potential inherent in optical systems is provided by scattering-assisted photonic reservoirs \cite{giganImagingComputingDisorder2022}. The basic operating principle in these systems is to encode the input data in the optical wavefront and propagate the structured field carrying the information through a random scatterer. The transmitted field is sampled by a CCD camera that measures the average intensity per pixel. The mathematical intuition behind this system is that the scattering medium effectively performs a complex-valued, random-matrix projection of the input wavefront (akin to a random connection layer in a neural-network-based reservoir). In contrast, the intensity detection of the output from the matrix projection provides the nonlinear transformation required for nonlinear processing tasks, particularly when considering CCD cameras operating outside of the linear operation region. Pivotal to this scheme are Spatial Light Modulators (SLMs) \cite{rafayelyanLargeScaleOpticalReservoir2020}, which serve as the translational interface between input data and its optical manifestation. Intuitively, wave mixing and nonlinear interactions can be understood as the two ingredients required to implement a controlled nonlinear transformation of the input data, in close analogy with the combination of linear connections and activation functions in multilayer perceptrons \cite{dongOpticalReservoirComputing2020,dongReservoirComputingMeets2020,pierangeliLivingOpticalRandom2020}. From a fundamental perspective, these photonic technologies offer the opportunity to explore the learning and processing dynamics of physical RC systems, offering a design space that naturally provides high-dimensional processing of input data. 

Despite their simplicity, scattering-based RC systems have demonstrated remarkable data-processing capabilities, particularly in the prediction of large-scale spatiotemporal series forecasting \cite{rafayelyanLargeScaleOpticalReservoir2020}, natural language processing \cite{valensiseLargescalePhotonicNatural2022}, and, more recently, in the hardware acceleration of Large Language Model (LLM) training via Direct Feedback Alignment \cite{wangOpticalTrainingLargescale2024}. As recognized in recent literature, a key advantage is provided by the scalability of this approach, as the time required by the matrix-vector multiplication taking place in the optical domain is independent of the data dimensionality and, therefore, of the matrix size (up to experimental latency). This means that, in principle, the combination of full-resolution SLMs and CCDs can perform algebraic manipulations at a rate limited only by the response time of the slowest electronic component (usually the SLM, operating at less than 60Hz for phase-only devices) and independently of the matrix size \cite{wangOpticalTrainingLargescale2024,wangOpticalNextGeneration2024}.  

Beyond technological and performance considerations, however, photonic reservoirs based on scattering-assisted wavefront shaping offer the opportunity to explore the learning and processing dynamics of physical reservoir computing (RC) systems, providing a design space that naturally supports high-dimensional processing of input data. In the present day, a fuller understanding of how \textit{physical} and \textit{informational} dynamics map is critical. Photonic reservoirs, hence, are a uniquely practical platform for a deeper understanding of the key principles and advantages provided by physical reservoir computing (RC) systems.

Among the most pressing questions is that of the fundamental role played by the \textit{encoding} of information into a physical substrate. The encoding, together with the readout scheme, is where the design space for physical RC is most open. This is particularly true for systems based on phase-only SLM systems, where the input data is encoded in a phase distribution impressed on an optical field. While it has been extensively recognized in recent works that the nonlinear transformation data~$\rightarrow$~optical phase effectively acts as a first nonlinear layer in the reservoir, a standard and intuitive recipe to perform this encoding dictates that the input data \textit{must} be mapped in the natural period of the optical waves $[0, 2\pi)$ \cite{Bauwens2022PhaseEncoding, Nguimdo2016PhaseSensitivity, wangOpticalNextGeneration2024, rafayelyanLargeScaleOpticalReservoir2020}.

The primary rationale is that, due to the periodicity of the phase encoding function $\phi(x)$ beyond this interval, mapping data outside of a $2\pi$ encoding region would lead to an immediate loss of bijectivity between the input data and the encoded waves. Following the nomenclature of signal analysis, we refer to this effect as \textit{phase wrapping}. Said differently, distinct data points mapped to angles spaced by $2\pi$ will result in the same (mathematically exact) encoded field. In the context of computation, this presents a potentially serious issue, as the \textit{separability} of outputs is no longer guaranteed. That is, two distinct input signals become indistinguishable at the output. This phenomenon extends beyond computing, and this effect is particularly pronounced in quantum tracking control, where it results in a bidirectional non-uniqueness between driving field and observable expectations \cite{mccaul_non-uniqueness_2022, masur_optical_2022, mccaul_optical_2021, mccaul_driven_2020, mccaul_controlling_2020}.   

For these reasons, phase encoding in both optical \cite{Bauwens2022PhaseEncoding} and quantum \cite{Govia2021QuantumPhaseRC} RC maps strictly to the $[0,2\pi)$ range. Particularly relevant to this work, the restriction to the unit circle is also a fundamental assumption in Random Fourier Features approaches in machine learning \cite{rahimiRandomFeaturesLargeScale2007}. While this domain restriction, by definition, ensures separability, it also forecloses a significant opportunity. Periodicity and phase wrapping are intrinsically \textit{nonlinear} features, and as such are dynamical resources that should not be left unharnessed. The challenge then is how to exploit the inherent periodicity of phase encoding \textit{without} running afoul of the loss of output separability it may induce. 

Here, we pursue this goal. To do so, we investigate the impact on computational and predictive capabilities of a scattering-assisted photonic RC when the phase encoding domain is extended by a \textit{wrapping factor} $\alpha$. Remarkably, we observe a counterintuitive result: phase wrapping becomes an asset for a range of data-processing tasks. At the heart of this result lies the recognition that phase encoding in a spatial light modulator naturally constructs the \textit{Fourier modes} of the information it encodes. The ability to manipulate the periodic patterns impressed by the phase encoding onto the data features \textit{across the dataset} through the combination of random masking and scattering-assisted wave mixing enables the generation of highly compressed nonlinear features spanning both the feature and sample dimensions of the data. 
Through formal analysis and numerical simulation, we demonstrate that phase wrapping corresponds to increasing the span of modes available for linear combination in the RC output layer. We demonstrate regression and classification tasks with significantly improved accuracy. Our results pave the way for a novel approach to designing photonic RC systems where non-bijective phase encoding is an essential ingredient.    

\section{Model}
\subsection{Photonic configuration\label{sec:setup}}
\begin{figure}[h!]
\centering
\includegraphics[width=\columnwidth]{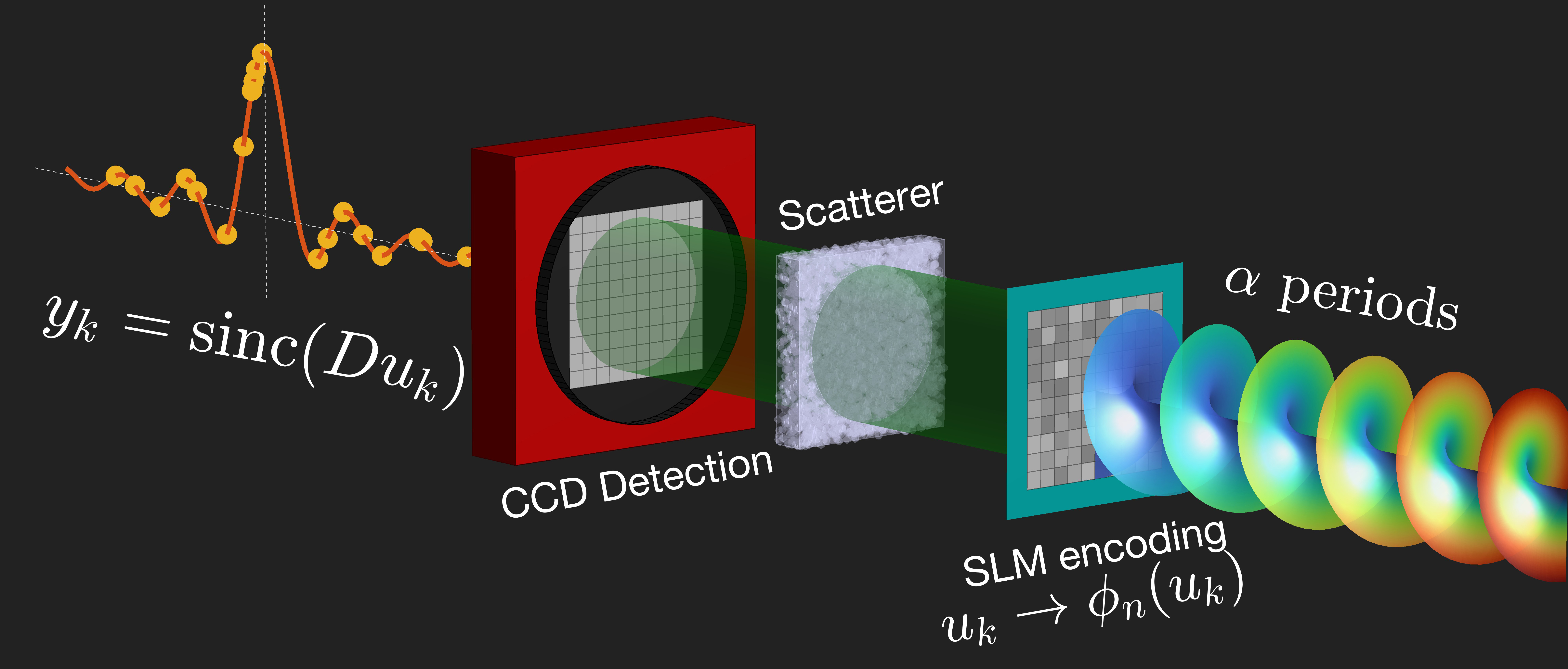}
\caption{\label{fig:setup}
    \textbf{System configuration}. Our photonic RC system is composed of a spatial light modulator (SLM), a random scatterer, and a CCD camera. The input data $u_k$ is encoded for each $n$-th pixel in the phase $\phi_n(u_k)$ of the optical field. The wrapping factor  $\alpha$ corresponds to a phase modulation spanning multiple $[0,2\pi)$ periods. The CCD images are collected and yield the predicted output (e.g., $y_k = \sinc(D u_k)$ via linear regression. 
}
\end{figure}
Our system is shown in Fig. \ref{fig:setup}. We encode input data $u_k$ in the phase front $\phi(\mb{x})$ of a spatial light modulator (SLM) composed of $N$ pixels. In our formalism, the transverse spatial coordinate $\mb{x}$ is discretized in a flattened (one-dimensional) spatial bins $x_n$, and the spatial distribution of the key quantities is discretized accordingly, e.g., $\phi(\mb{x}_n)\equiv\phi_n$. 
The optical field, shaped by the SLM, is imaged on the input surface of a scattering medium, and we denote the field immediately before and after the scatterer as ${E}^-(x_n)$ and $E^+(x)$, respectively. Upon phase-modulation, the input field reads:
\begin{equation}
    \label{eq:fie0}
    E^-_n(u_k) = {\exp\left[\im \phi_n(u_k) \right]}/{\sqrt{N}}, 
\end{equation}
and we model the propagation through the scattering system through a scattering matrix formalism: 
\begin{equation}
    \label{eq:fie1}
    E^+_m(u_k) = \sum_n T_{mn} E^-n(u_k).
\end{equation}
The scattering matrix elements $T_{mn}$ connect the $n$-th input pixel with the $m$-th output pixel, and are defined as:
\begin{equation}
    \label{eq:tm0}
    T_{mn} = \frac{a_{mn} + \im b_{mn}}{\sqrt{N}},
\end{equation}
In Eq. \eqref{eq:tm0}, $a_mn$ and $b_mn$ are sampled from a gaussian distribution with mean $\mu=0$ and standard deviation $\sigma = 1/2$ \cite{goodmanStatisticalOptics2015, kumarDeterministicTerahertzWave2022a}. The output field is imaged on the sensor of a CCD camera, leading to the intensity detection: 
\begin{equation}
    \label{eq:ccd0}
    I_m(u_k) = h_{CCD}(E^+_m), 
\end{equation}
where $h_{CCD}$ is the readout function of the camera. Without loss of generality, we initially assume that the camera has been calibrated to ensure the dynamic range of the field output lies in the linear response region of the CCD sensors, i.e., $h_{CCD}(E^+_m)=\modd{E^+_m}^2$. 

To verify the robustness of our results against realistic experimental conditions, we also extended our model to include the effects of finite bit depth in the CCD and SLM, as well as the presence of shot noise \cite{heOptimalQuantizationAmplitude2021, janesickScientificChargeCoupledDevices2001}. We model a bit-depth-limited SLM by quantizing the phase in $b_{SLM}$ bits, i.e.:
\begin{equation}
    \label{eq:bit0}
    \phi_n(u_k) \rightarrow \Phi_n(u_k) = \floor{\frac{\phi_n(u_k)}{\Delta}+\frac{1}{2}}\Delta.
\end{equation}
where $\Delta = 2\pi / 2^{b_{SLM}}$ and $\floor{\cdots}$ is the floor function. This quantization leads to a finite number of phase values that the SLM can encode.
To include the effects of bit-depth-limited CCD detection, we first normalize the intensity $I_m(u_k)$ as follows: 
\begin{equation}
    \label{eq:bit1}
    \mc{I}_m(\mb{x}) = \frac{I_m(u_k)}{I_{SAT}}g + b ,
\end{equation}
where $g$ and $b$ are tunable gain and offset parameters, and $I_M$ is the saturation intensity acting as a normalization factor and fixed across all samples and pixels. To model the effect of shot-noise, we sample from a Poisson statistical distribution centered at $\mc{I}_m(\mb{x})$:
\begin{equation}
    \label{eq:bit2}
    P_m(u_k) \sim \text{Poisson}\left(I_{MAX}\mc{I}_m(\mb{x})\right).
\end{equation}
where $I_{MAX} = 2^{b_{CCD}}-1$ is the maximum depth. Finally, the resulting intensity is clipped and quantized in $b_{CCD}$ bits, i.e.:
\begin{equation}
    \label{eq:bit3}
    {I}_m(u_k) = {\floor{\min\left\{I_{MAX},{P}_m(u_k)\right\}+1/2}}/{2^{b_{CCD}}}.
\end{equation}
In our simulations, we set $b_{SLM}=8$ and $b_{CCD}=16$, corresponding to standard SLM and CCD scientific cameras. 

\subsection{Reservoir computing pipeline\label{sec:rc}}
Consider a given dataset $\{\mb{U}, \mb{Y}\}$, where $\mb U=[u_1, u_2, \cdots, u_k]$ is the input data and $ \mb{Y}=[y_1, y_2, \cdots, y_k]$ is the output (ground-truth) data.  The number of samples in the dataset is denoted by $K$, and the dataset is split into training $\{\mb{U}_{train}, \mb{Y}_{train}\}$ and $\{\mb{U}_{test}, \mb{Y}_{test}\}$, with sizes $K_{train}$ and $K_{test}$, respectively. In the following, we retain a one-dimensional formulation for conciseness; however, the extension to multi-dimensional input and output data (with dimensions $d_u$ and $d_y$, respectively) is straightforward. In the following, we retain a one-dimensional formulation for conciseness; however, the extension to multi-dimensional input and output data (with dimensions $d_u$ and $d_y$, respectively) is straightforward. The dataset processing on our optical system consists of three steps: encoding, training, and testing. 

First, we encode each training input data point $u_k$ on the phase of the SLM:
\begin{equation}
    \label{eq:ph0}
    \phi_n(u_k) =   2\pi\alpha g_n(u_k)\mod{2\pi},
\end{equation} 
where $\alpha$ is the scaling factor introducing the phase wrapping and $g_n(u_k)$ is an embedding function that maps the input data $u_k$ to a phase value $\phi_n(u_k)$ for the $n$th pixel of the SLM. Before encoding, the input data is normalised to the interval $[0,1]$. Given $\phi_n(u_k)$ is a phase with underlying periodicity $ 2\pi$, it is generally expected that the best performance is achieved when $\alpha \leq 1$. In our case, conversely, we explicitly consider arbitrary wrapping factors and apply a $\mod{2\pi}$ operation on the stretched phase. While the choice of embedding function $g_n(u_k)$ is arbitrary and generally task-dependent, we consider linear encodings defined as:
\begin{equation}
    \label{eq:enc0}
    g_n(u_k) = G_n u_k,
\end{equation}
where $\mb{G} = \{G_n\}$ is a fixed $N$-dimensional mask vector (or a $N\times d_u$ matrix for multi-dimensional data). The mask vector $G_n$ can be chosen to be linearly spaced or randomly distributed, leading to different wrapping behaviors. 

In the training stage, we project the patterned input fields on the scattering medium and collect the (flattened) CCD images in an $(N \times K_{train})$ matrix $\mb{R}$. Note that the ground-truth data $y_k$ is not part of the training process. 
Once the training is completed, we perform a Ridge regression of the reservoir matrix $\mb{R}$ and estimate the best-guess of the readout weights as:
\begin{equation}
    \label{eq:rid0}
    \mb{W}_{out} \mb{Y}_{train} \mb{R}^T \left( \mb{R}^T\mb{R} + \lambda \mb{I} \right)^{-1} ,
\end{equation}
where $\lambda$ is the ridge regularization coefficient \cite{hastie2009elements,hastieRidgeRegularizationEssential2020} in such a way that our predicted output reads as follows: 
\begin{equation}
    \label{eq:rid1}
    \hat{\mb{Y}}_{train} = \mb{W}_{out} \mb{R}.
\end{equation}

Finally, in the testing stage, we encode and project each data point in the test dataset, collect the CCD images in a $N\times K_{test}$ test matrix $\mb{R}_{test}$, and calculate our predicted outputs as
\begin{equation}
    \label{eq:rid2}
    \hat{\mb{Y}}_{test} = \mb{W}_{out} \mb{R_{test}}
\end{equation}
As figures of merit for the training and test datasets, we compute the Normalized Mean Squared Error (NMSE): 
\begin{equation}
\label{eq:err0}
    NMSE[\mb{Y},\hat{\mb{Y}}] =
    \dfrac{\sum_k \left(y_k - \hat{y}_k \right)^2}
    {\sum_k \left(y_k - \bar{Y} \right)^2}
\end{equation}
where $\bar{{Y}}=\sum_k y_k / K $ is the mean of the ground-truth values. We note here that our system is purely feed-forward, and formally similar to an Extreme Learning Machine (ELM) rather than a time-dependent RC system. Since we only focus on time-independent tasks, and this scheme can be extended to spatiotemporal forecasting systems through output feedback \cite{Dong2018ScalingRC}, we will retain the term reservoir for the sake of generality.  

Across all our simulations, the Ridge regression was performed using a Leave-One-Out Cross-Validation (5-fold) routine (\verb+RidgeCV+ from the \verb+scikit-learn+ library). Unless specified otherwise, we considered a fixed ridge coefficient $\lambda = 10^{- 6}$. For classification tasks, we also define classification accuracy as the percentage of correctly classified samples in the training and test datasets, obtained via a winner-takes-all readout where the predicted class $\hat{y}_k$ is the class with the maximum output value. For visualization purposes, we also computed the $F_1$ score using a weighted average. Unless specified otherwise, the classification is generally based on a one-hot encoding of the class labels \cite{hastie2009elements}. 

\section{Separability considerations in phase-wrapped encoding \label{sec:encoding}}
Having established the experimental setup and its formal representation, we now turn to motivating the key novelty of our work: the expressivity-boosting role of the phase wrapping factor $\alpha$. Here, the term expressivity refers to the ability of the reservoir to generate a rich set of features that can be linearly combined to approximate the target function. 

In supervised learning tasks, two key concepts are generally considered essential for successful data-processing: \textit{separability} and \textit{nonlinearity} \cite{hastie2009elements}. Separability refers to the ability to distinguish between different data points, while nonlinearity enables the construction of complex features of the input data (e.g., image transformations, decision boundaries) necessary to identify the connection between input and output data successfully. The nonlinear features, then, enable fine-tuning the model on the task at hand. In RC systems, these two concepts are intimately linked to the choice of encoding function and the subsequent processing of the encoded data, with the added complication that the target function is generally not known during the training process. Consequently, a "successful" reservoir must generate a wide set of nonlinear features ensuring the right balance of separability and \textit{expressivity} to capture the underlying data structure \cite{daleRoleStructureComplexity2019}. 

If both $u_k, G_n \in [0,1]$, then $\alpha$ directly controls the degree of possible phase wrapping latent in this encoding. The standard condition $\alpha < 1$ trivially satisfies the \textit{separability} condition, where if 
\begin{equation}
\label{eq:eq:sep_cond}
    u_k \neq x_{k'} \implies I_m(u_k) \neq I_m(x_{k'})
\end{equation}
then 
\begin{equation}
   \exists\, m \in \{1, \dots, N\} \;\text{such that}\; I_m(u_k) \neq I_m(x_{k'}).
\end{equation}
That is, two distinct inputs are considered separable if they yield distinct intensity values in at least one detector pixel. For $\alpha < 1$, the map $u_k \rightarrow I_m(u_k)$ is trivially injective across all $m$, since the maximum phase range across the entire SLM remains strictly within $[0,2\pi)$. In this regime, no two inputs can produce overlapping phase wraps; thus, the reservoir exhibits guaranteed separability.

On the contrary, any $\alpha > 1$ admits the possibility of a non-bijective mapping from data to single pixel input features. It is worth pausing here to clarify a subtle but crucial point regarding the physics underlying phase manipulation. While the optical field is itself periodic, the phase introduced by our encoding does not directly correspond to any variation in a physical parameter such as wavelength, momentum, or spatial frequency. Instead, in our case, the phase shift corresponds to a synthetic structure imposed on the data by the joint choice of $g_n$ and $\alpha$. This reframing emphasizes that the periodicity we aim to exploit is not that of the optical carrier, but that of the data embedding itself. 

\begin{figure*}[htbp!]
\centering
\includegraphics[width=\textwidth]{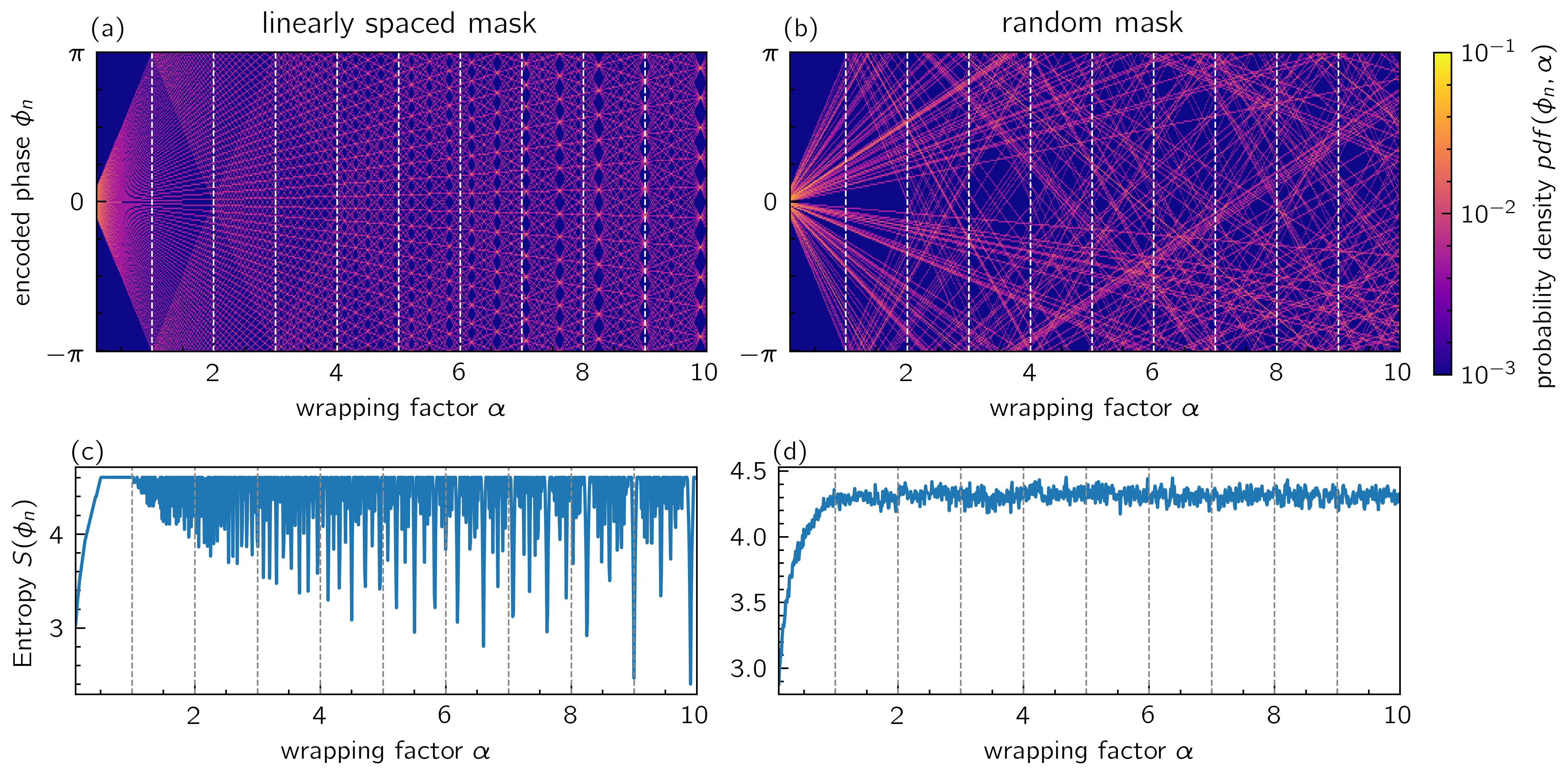}
\caption{\label{fig:pdf}
    \textbf{Probability analysis of angular encoding}. (a, b) Probability density functions (in $\log_{10}$ scale) of the angular frequencies generated by an equispaced mask (panel a) and a uniform randomly distributed mask (panel b) as a function of the wrapping factor $\alpha$. The vertical dashed lines are visual guides for the multiple encoding function periods. (c,d) Shannon Entropy as a function of $\alpha$ for the probability distributions associated with equispaced mask (panel c) and random mask (panel d). Here, we considered $N=100$, and the probability density functions were computed over 200 bins in the interval $[0, 2\pi]$.
}
\end{figure*}

To understand the effect of phase wrapping in this context, we first analyze the statistical probabilities of the phases generated by our encoding function $\phi_n(u_k)$ as a function of the wrapping factor $\alpha$. To do so, we consider a set of $N$ pixels and trace the probability density function (PDF) of the angular frequencies generated by the phase encoding function $\phi_n(u_k)$ for two different choices of the mask vector $\mb{G}$: an equispaced mask and a uniformly randomly distributed mask. The PDF is computed over 200 bins in the interval $[0, 2\pi)$ and normalized to ensure that it integrates to one. 

The results of the probability analysis are shown in Figure \ref{fig:pdf}. In panel (a), we show the PDF for an equispaced mask, while in panel (b), we show the PDF for a uniformly randomly distributed mask. The vertical dashed lines in both panels are visual guides for the multiple encoding function periods. As clearly seen in panel (a), once the wrapping factor $\alpha$ exceeds one, the PDF becomes increasingly concentrated around the multiples of $2\pi$, resulting in a loss of separability and periodic patterns in the encoded phases. This is because the phase encoding function $\phi_n(u_k)$ becomes non-bijective, and distinct data points lead to the (mathematically exact) same phase value. To illustrate this, we also compute the Shannon entropy of the PDF as a function of $\alpha$, shown in panel (c). As expected, the entropy decreases as $\alpha$ increases, indicating a loss of information and separability across $\alpha$. 

In contrast, for the uniformly randomly distributed mask (panel b), the PDF remains more uniformly distributed across the interval $[0, 2\pi)$, even for larger values of $\alpha$. This result suggests that, at the encoding stage, the random mask can mitigate the loss of separability associated with phase wrapping, as further confirmed by the Shannon entropy shown in panel (d), which remains relatively constant across $\alpha$. The random mask preserves the information content of the encoded phases, even when the wrapping factor $\alpha$ exceeds one. With regard to separability, for truly random continuous $G_n$, the equation $I_m(u_k) = I_m(u_{k'})$ with $u_k \neq u_{k'}$ has no deterministic solutions – it would require an accidental cancellation or alignment of random phasors.

\section{Results: Nonlinear $\sinc$ Regression \label{sec:sinc}}
\begin{figure*}[htbp!]
\centering
\includegraphics[width=\textwidth]{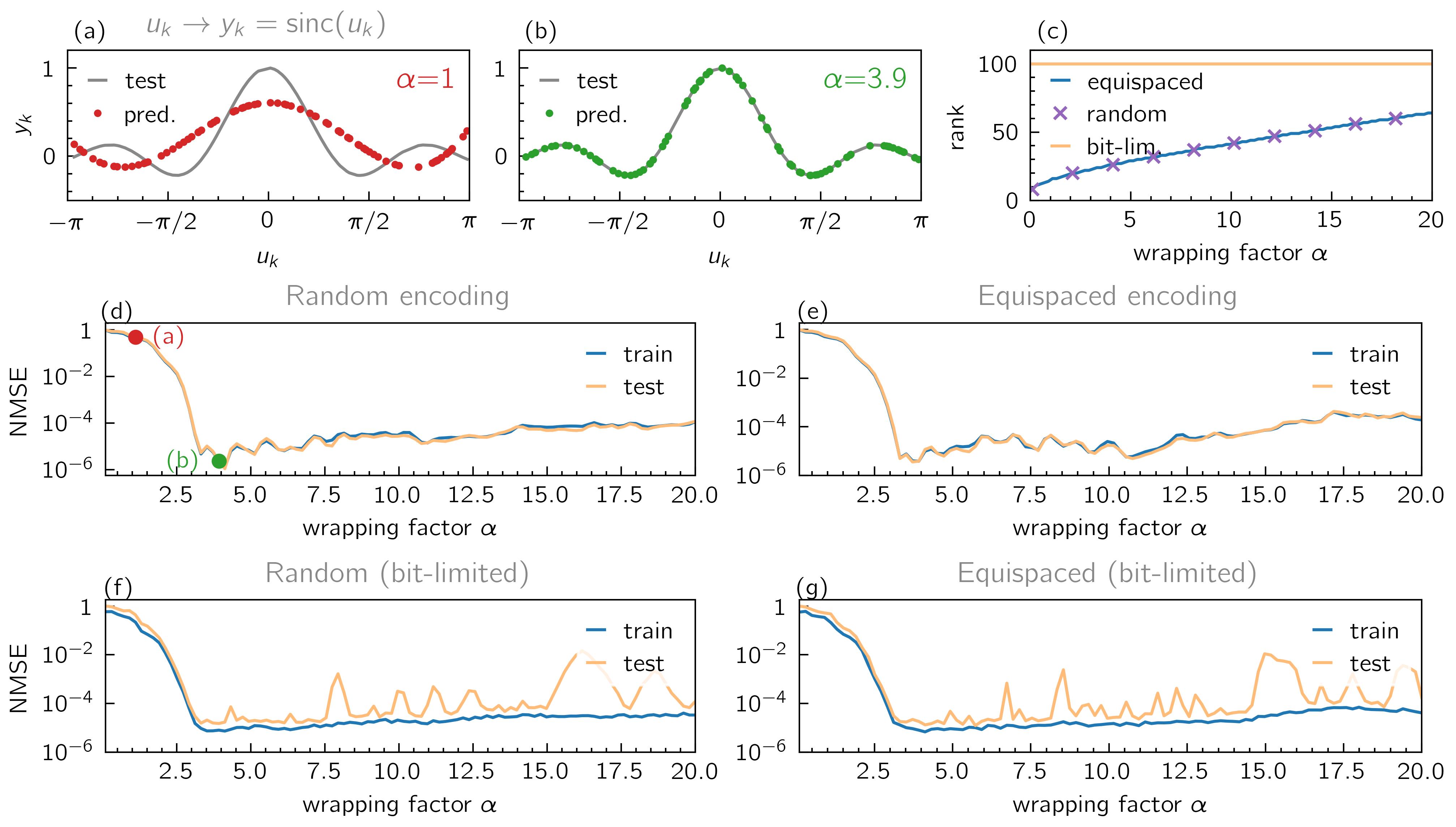}
\caption{\label{fig:sinc}
    \textbf{Sinc Regression Results}. (a) Predicted output $\hat{y}_k$ (dots) and ground truth function $y_k$ for $\alpha=1$ (standard bijective encoding). (b) Same as panel a, but for an optimal value of $\alpha$=3.9. (c) Reservoir matrix ranks for equispaced (solid line) and random encodings (dots, shown one every 5 data points). The yellow line corresponds to the bit-depth-limited scenario. (d,e) Train (blue line) and test (yellow line) NMSE as a function of the wrapping factor $\alpha$. Panel d corresponds to random $G_n$ masks, while panel e is obtained via equispaced encoding. (f,g) Same as panels d and e, but considering a bit-depth-limited configuration. Simulation parameters: $N=100$, $K_{train}=500$, $K_{test}=200$.    
}
\end{figure*}

Equipped with this understanding, we now turn to the impact of phase wrapping on the separability and nonlinearity of the reservoir output by considering the impact of the phase wrapping on a standard universal nonlinear regression task \cite{marcucciTheoryNeuromorphicComputing2020,teginScalableOpticalLearning2021,RCMMFGain}. Specifically, we consider the fitting of the nonlinear function 
\begin{equation}
    \label{eq:sinc0}
    u_k \longrightarrow y_k = \sinc(D u_k) = \frac{\sin(\pi D u_k)}{\pi D u_k},
\end{equation}
where $D$ is a constant and the input data $u_k$ is randomly sampled in $[-\pi, \pi]$.  Despite its simplicity and ubiquity, this function is known to be particularly challenging for photonic RC systems, as it requires a high degree of nonlinearity and sensitivity to small variations in the input data \cite{huangUniversalApproximationUsing2006}. Following standard approaches, we encode the input data-points in the optical domain via an equispaced random lifting $\mb{G} = \{G_n\}$ where a fixed random vector of size $N\times 1$ is uniformly distributed in $[0,1]$. Critically, this scheme maps a single datum into $N$ phases.

Figure \ref{fig:sinc}(a) illustrates the results for our regression task when considering this scenario with $N=100$. As evident from the figure, the regression fails, resulting in an NMSE = 0.48. Quite interestingly, the prediction is not entirely random, but the predicted function resembles a $\sinc$ function defined on a smaller restricted domain. An unexpected result, however, occurs when $\alpha$ is increased beyond the natural value, as shown in Figure \ref{fig:sinc}(b), where $\alpha = 3.9$. In this scenario, the ridge regression of the reservoir output leads to a remarkably accurate fitting of the $\sinc$ function, with an MSE=2.34$\times 10^{-6}$. 
This result suggests that the phase wrapping induced by the larger $\alpha$ value enables the reservoir to effectively sample a broader range of nonlinear features, resulting in a more accurate reconstruction of the target function. To better estimate the effect of the embedding scaling factor, we performed a series of simulations with $\alpha$ values ranging in $[0.1, 20]$. The results are shown in Figure \ref{fig:sinc}(d), where we observe a clear trend of decreasing MSE error as $\alpha$ increases, up to a value of $\alpha~3.7$ (NMSE=2.34$\times 10^{-6}$). Beyond this point, the MSE error starts to increase again, indicating that there is an optimal value of $\alpha$ that maximises the reservoir expressivity for this task. Quite significantly, we verified that the decrease in error is not associated with the algebraic properties of the reservoir matrix, as confirmed by the matrix ranks included in Fig. \ref{fig:sinc}(c). The lack of any correspondence between regression minima and rank suggests the reservoir is generating a compressed representation of the input data. In the bit-depth-limited scenario (\ref{fig:sinc}(c), yellow line), the presence of shot-noise automatically makes the matrix full-rank. 

Perhaps more surprisingly, however, the results in Figure \ref{fig:sinc}(d) correspond to an equispace encoding scheme, where the embedding vector $\mb{G}$ is uniformly distributed in $[0,1]$. Even in this scenario, we observe a similar trend to the random embedding in Fig. \ref{fig:sinc}(d), with an optimal value of $\alpha$ very close to the equispaced case ($\alpha=3.72$, NMSE=3.65$\times 10^{-6}$). A similar trend is observed also for bit-depth-limited conditions, as shown in Figure \ref{fig:sinc}(f)-(g), confirming that the expressivity enhancement should be observable in realistic experimental conditions.

These results are particularly significant, as they demonstrate that the phase wrapping induced by the larger $\alpha$ value enables the reservoir to effectively sample a broader range of nonlinear features, resulting in a more accurate reconstruction of the target function that circumvents the loss of separability associated with the phase wrapping. This is a key result of our work, as it shows that phase wrapping can be a powerful tool for enhancing the expressivity of photonic RC systems, even in the presence of non-bijective mappings.
 
\section{Nonlinear feature engineering via Random Fourier components \label{sec:fourier}}
\begin{figure*}[htbp!]
\centering
\includegraphics[width=\textwidth]{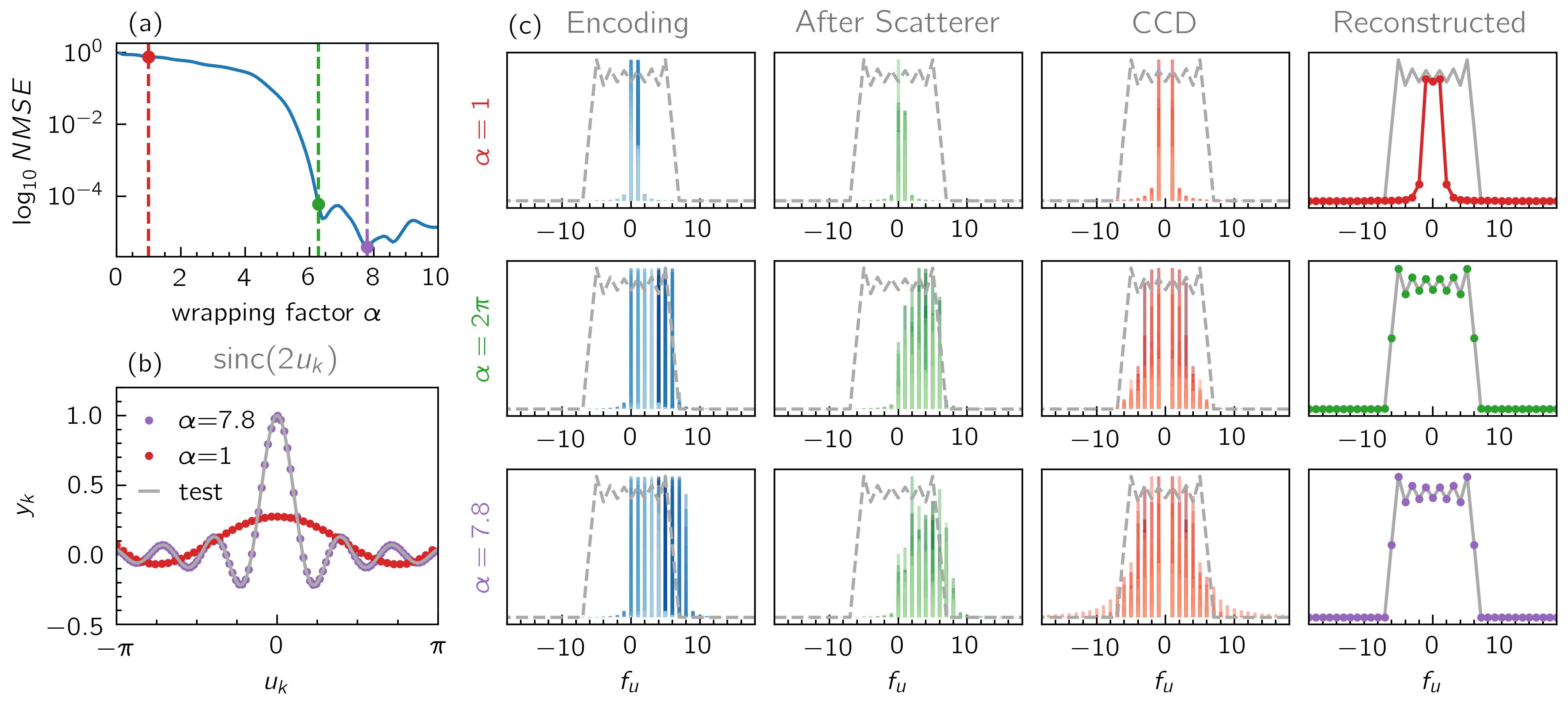}
\caption{\label{fig:freq}
    \textbf{Sinc regression as a spectral reconstruction problem.}. (a) NMSE regression error as a function of the wrapping factor $\alpha$ for the $\sinc$ regression task with $N=100$ pixels and $K_{train}=500$. (b) Same as panel a, but for a bit-depth-limited SLM and CCD configuration. (c) Spectral content of the encoding (first column), transmitted field (second column), ccd pixels (third column), and predicted output (fourth column) for different values of $\alpha$. The black dashed line indicates the spectral content of the target function. In the CCD plots (third column), the average intensity per pixel was subtracted to highlight the spectrum at non-zero frequencies (leading to a null zero frequency component). 
}
\end{figure*}
To understand the mechanism by which phase wrapping enhances the expressivity of the reservoir, we must first consider the nature of the features generated by the SLM. In the context of our system, we can think of the SLM as a mechanism that encodes the input data into a set of spectral modes, each corresponding to a pixel on the SLM. Consequently, the phase encoding function $\phi_n(u_k)$ effectively defines a set of \textit{synthetic Fourier modes} that are sampled from the input data. In this regard, the phase encoding can be seen as a physical realisation of the Random Fourier Features models \cite{rahimiRandomFeaturesLargeScale2007}. It is this representational geometry — defined entirely by the embedding — that enables phase wrapping to enhance rather than degrade expressivity. To detail the mechanism by which this occurs, it is helpful to maintain this perspective: physical encoding by the SLM corresponds to the construction of a set of spectral modes, with frequencies determined by the embedding function -or more precisely, the linear mask $G_n$. 

To illustrate this point, we considered again the $\sinc$ regression task, but this time we considered a set of ordered equispace input data $u_k = 2\pi(k/K)-\pi$ where $k=1, \dots, K$ and $K=500$ is the total number of samples. Similarly to the case in section \ref{sec:sinc}, the input data is encoded in the phase of the SLM via an uniformly random mask $G_n$ and $N=100$ is the number of pixels. The results are shown in Fig. \ref{fig:freq}, with \ref{fig:freq}(a) illustrating the NMSE regression error for the task $u_k \rightarrow y_k = \sinc{2 u_k}$. Even in this scenario, the system fails to reconstruct the output for $\alpha=1$ (NMSE=0.76, red dots in Fig. \ref{fig:freq}(b)), but is able to achieve an NMSE=3.87$\times10^{-6}$ for $\alpha=7.8$, corresponding to the purple dots in Fig. \ref{fig:freq}(b).  

Equipped with equispaced input points, we can explicitly compute the frequencies generated by the phase encoding via FFT of the information carried by each input and output pixel as it propagates through the system. The three rows correspond to the three dots in Fig. \ref{fig:freq}(a), corresponding to $\alpha=1$ (red), $\alpha=2 \pi$ (blue), and $\alpha=7.8$ (purple). The first row corresponds to the case where the phase encoding is bijective, while the second and third rows correspond to the cases where the phase encoding is non-bijective and phase wrapping boosts expressivity. The results are shown in Fig. \ref{fig:freq}(c), and will be discussed in detail below. In all panels, the black dashed line indicates the spectral content of the target function, i.e., the $\sinc{2u}$ function.

At the encoding stage (first column), we observe that the phase encoding generates a set of spectral modes with frequencies determined by the mask $G_n$ and the wrapping factor $\alpha$. For $\alpha=1$, the spectral content is limited to a small number of modes. As $\alpha$ increases, the spectral content becomes richer, and the case $\alpha=2\pi$ essentially corresponds to frequencies covering the $\rect(f/2)$ function associated with the Fourier transform of the target. The case $\alpha=7.8$ corresponds to a further increase in the spectral content, with frequencies covering a wider range of the Fourier spectrum and compensating for the truncated input domain $u_k \in [-\pi,\pi]$. These results provide a first confirmation that a successful regression is driven by the ability of the reservoir to sample a rich set of spectral modes, which in turn enables the reconstruction of the target function. It is through the subsequent action of the scatterer and CCD detection, however, that the features generated by this mode decomposition can be exploited. 

As they propagate through the system the synthetic frequencies injected by the encoding undergo two main transformations: first, they are mixed by the scattering matrix $T_{mn}$, whose role is to distribute the encoded synthetic frequency modes across the output pixels of the CCD camera. This effect is visible in Fig. \ref{fig:freq}(c), second column, where we show the spectral content of the transmitted field.  

\begin{figure}[htbp!]
\centering
\includegraphics[width=\columnwidth]{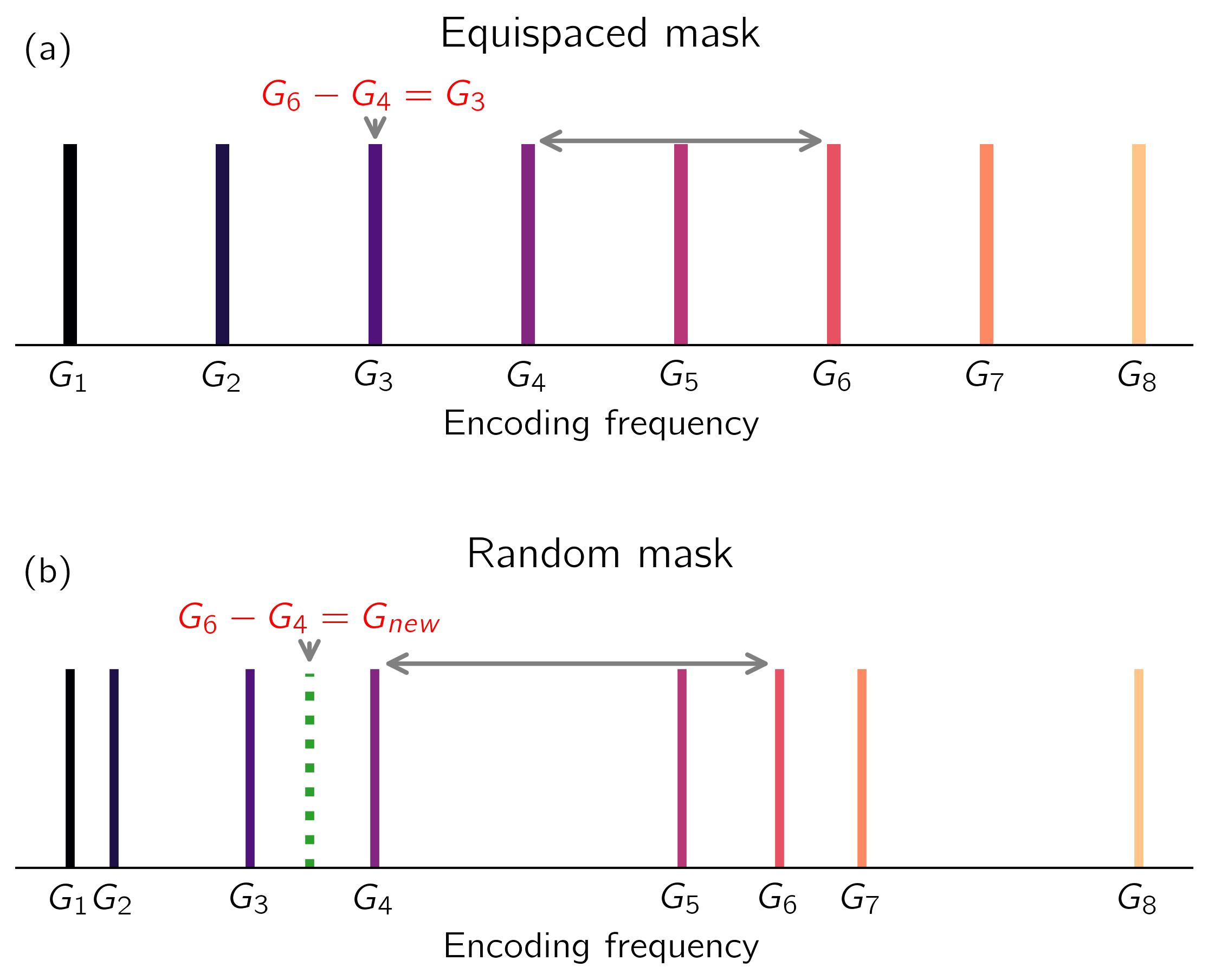}
\caption{\label{fig:freqmix}
    \textbf{Nonlinear frequency mixing}. Illustration of the generation of existing (panel a, equispaced masking) and new frequencies (panel b, random masking) from the combination of encoding~$\rightarrow$~scatterer~$\rightarrow$~intensity detection. In the equispaced case, the generated frequencies are a subset of the original ones, while in the random case, new frequencies are generated via nonlinear mixing.
}
\end{figure}

A more careful analysis of Eqs. \eqref{eq:fie0}-\eqref{eq:ccd0}, however, sheds light on a key nonlinear mechanism originating from the combination of phase encoding, random scattering, and intensity detection. By expanding the intensity readout in Eq. \eqref{eq:ccd0} for a linear intensity response $I=\modd{E}^2$, we can rewrite the output intensity as:
\begin{equation}
    \label{eq:mix}
  I_m(u_k)=|E^+_m(u_k)|^2=\sum_{np} T_{mn} T_{mp}^* \exp{[2 \pi i \alpha (G_n - G_p^*)u_k]}  
\end{equation}
Eq. \eqref{eq:mix} highlights a key feature of the scattering-assisted reservoirs: while the phase encoding controls the number and range of accessible frequencies, the combination of scattering-induced mixing and intensity detection leads to the generation of new synthetic frequencies via frequency-difference nonlinear mixing, as illustrated in Fig. \ref{fig:freqmix}. The increase in available frequencies is indeed a key mechanism enhancing the expressivity of our reservoir, as it allows the system to sample a wider range of Fourier modes and thus better reconstruct the target function even in the presence of a small number of pixels. This mechanism is in line with the quantum Extreme Learning Machine (ELM) formalism, wherein for a fixed $\alpha$, the difference or spacing between the eigenvalues of the Hermitian generator $H_j$ determines the number of accessible frequencies and has been demonstrated to increase the expressivity of small-scale quantum reservoirs \cite{Xiong2025QELM, PhysRevA.103.032430}. 

The spectral content associated with the CCD pixels is shown in the third column of Fig. \ref{fig:freq}(c). As expected, the spectral content of the CCD pixels is a linear combination of the spectral content of the transmitted field, with each pixel corresponding to a different linear combination of the input modes. The intensity detection, moreover, fills the spectrum with a symmetric set of negative frequencies, enabling the real-valued regression. 

Finally, the fourth column shows the spectral content of the predicted output, which is obtained by applying a ridge regression to the CCD pixels. As evident from the figure, the predicted output closely matches the spectral content of the target function, confirming that the reservoir can effectively sample and reconstruct the target function via phase wrapping and nonlinear frequency mixing. The result for $\alpha=1$ (first row, red dots) is worth noting, as the reduced frequency content of the reservoir leads to a narrower predicted spectra corresponding to a $\sinc$ function defined in a broader domain (cf. Fig \ref{fig:sinc}(a) and Fig. \ref{fig:freq}(b)). This is a key result of our work, as it shows that the phase wrapping and nonlinear frequency mixing can be used to enhance the expressivity of photonic RC systems, even in the presence of non-bijective mappings.

\section{Results: spiral classification dataset}
\begin{figure*}[htbp!]
\centering
\includegraphics[width=\textwidth]{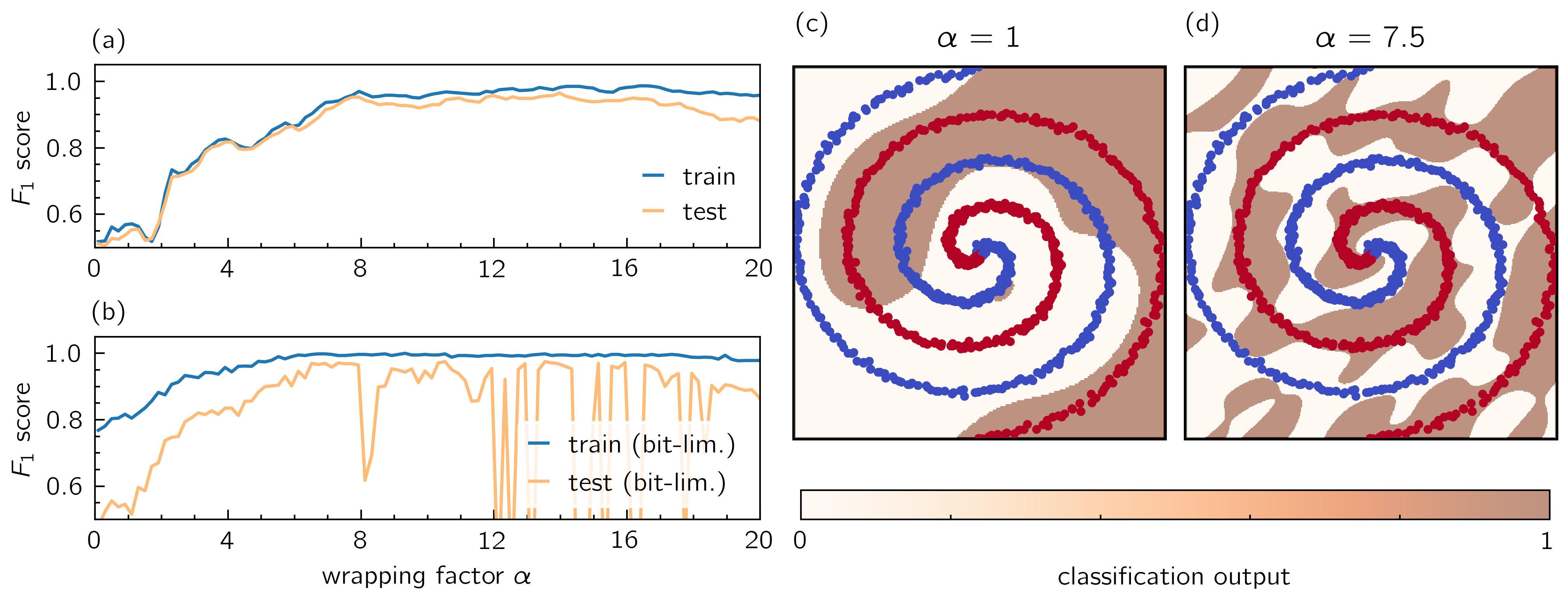}
\caption{\label{fig:spiral}
    \textbf{Spiral Classificaiton results}.
    (a) Classification accuracy (expressed by the $F_1$ score) for the spiral dataset as a function of the wrapping factor $\alpha$. The blue line represents the training dataset, while the yellow line represents the testing dataset. (b) Same as panel a, but considering a bit-depth-limited SLM and CCD configuration. (c) Spiral dataset in the $(u_{1,k},u_{2,k})$ plane, where the two spirals are superimposed to the decision boundaries of the reservoir for $\alpha=1$. (d) Same as panel c, but for an optimal value of $\alpha=3.9$. Simulation parameters: $N=100$, $K_{train}=600$, $K_{test}=400$. The dataset is generated with a winding number $\chi=2$ and a Gaussian noise of variance $\eta=0.05$. The decision boundaries are obtained via a winner-takes-all readout of the reservoir output.
}
\end{figure*}
While nonlinear regression tasks can help assess the expressivity of the system, they do not necessarily reveal the nonlinear separation capabilities of the reservoir. To address this point, we consider the spiral classification dataset \cite{saeedNonlinearInferenceCapacity2025a,RCMMFGain}. The dataset consists of two spirals in the $(u_{1,k},u_{2,k})$ plane, each containing 500 points, and is generated by the following equation system:
\begin{equation}
    \label{eq:spiral}
    \begin{cases}
    u_{1,k} = t_k \left( \sin(2\pi \chi t_k +c_k\pi) +\eta_k \right)\\
    u_{2,k} = t_k \left( \cos(2\pi \chi t_k +c_k\pi) +\eta_k \right)
    \end{cases}
\end{equation}
where $c_k=0$ for the first spiral and $c_k=1$ for the second spiral, $t_k$ is a parameter that varies from 0 to 1, $\chi$ is a winding number controlling the number of revolutions of the spiral branches, and $\eta_k$ is an uncorrelated Gaussian random variable (noise) with zero mean and variance $\eta$. 
The dataset is split into a training set with $M_{\rm train}=700$ and a testing set with $M_{\rm test}=300$. The target output for the binary classification is defined as the scalar $c_k$.
In this case, the embedding function is defined as follows: 
\begin{equation}
    \label{eq:emb1}
    \mb{\theta}_k(\mb{x}) = \mb{G}(\mb{x}) \left[\mb{S}(\mb{x}) u_{1,k} + [1-\mb{S}(\mb{x})] u_{2,k}\right],
\end{equation}
where $\mb{G}(\mb{x})$ is a random uniform vector in $[0,1]$. 
Each input data component $u_{1,k},u_{2,k}$ is normalised to the interval $[0,1]$, and $\mb{S}(\mb{x})$ is a random binary mask (with complement $[1-\mb{S}(\mb{x})]$). In practice, the embedding in Eq. \eqref{eq:emb1} randomly assigns $u_{1,k}$ and $u_{2,k}$ to distinct pixels on the SLM, and the resulting pattern is then randomly weighted as in Eq. \eqref{eq:ph0}.  

As an exemplary dataset, we considered a winding number $\chi=2$ and a Gaussian noise of variance $\eta=0.05$. 
In Fig. \ref{fig:spiral}(a), we report the classification accuracy for the binary classification $(u_{1,k},u_{2,k})\rightarrow c_k\in\{0,1\}$. The results demonstrate that the system achieves a classification accuracy close to 100\% for both the training and testing datasets across regions, aligning with the nonlinear regression scenario described in Sec.~\ref{sec:sinc}. As in the $\sinc$ regression task, a similar trend is observed in bit-depth-limited conditions, as shown in Fig. \ref{fig:spiral}b.

The dataset considered for this task is shown in Fig. \ref{fig:spiral}(c) and (d), where the two spirals are superimposed on the decision boundaries of the reservoir. In the bijective domain scenario ($\alpha=1$), the two spirals are not separable, leading to a classification accuracy below 50\% (random guess). As $\alpha$ increases, the reservoir output becomes more sensitive to slight variations in the input data, leading to a better separation of the two spirals. The decision boundary for the best configuration is shown in Fig. \ref{fig:spiral}(d) and indicates that the reservoir has learned to classify the two spirals correctly.

\section{Discussion and Conclusions}

In this work, we have investigated the role of phase encoding beyond its canonical $[0,2\pi)$ interval. We have shown that a significant enhancement of reservoir performance occurs when the encoding's wrapping factor $\alpha > 1$. While this result is counterintuitive at first, our analysis reveals a readily identifiable mechanism in the form of the encoding's embedding mask $G_n$. Though its function is to lift the data to the dimensionality of the reservoir's physical components, the effect of this procedure is to instantiate the data in the form of its Fourier modes, sampled at frequencies corresponding to $G_n$. Crucially, the combination of random scattering and CCD intensity detection induces a nonlinear mixing mechanism among these modes, which enhances the spectral content of the reservoir's nonlinear features. The combination of these elements results in the wrapping factor acting as a control for the bandwidth of Fourier modes available at the RC output layer. The predicted wrapping enhancement was then demonstrated both numerically and experimentally, with dramatic performance enhancements observed as a function of $\alpha$. In particular, we have shown that the reservoir can achieve high accuracy in both tasks, even under conditions with a limited number of pixels and bit-depth-limited conditions.

The expressivity enhancement effect driven by phase wrapping is precisely the type of behavior desired in reservoir computing: the construction of a high-dimensional, nonlinear feature space in which inputs become more easily separable by a linear readout while retaining all possible sample-to-sample correlations required to build arbitrary targets. Rather than harming expressivity, phase wrapping under $\alpha > 1$ produces a controlled explosion in mode interactions, turning linear embeddings into structured nonlinear kernels. The key is that this nonlinearity arises not from engineered activation functions, but from the internal geometry of the physical encoding and interference process itself. In this regard, a possible connection with standard RC lies in how the complex nonlinear system expresses the required features: in time-dependent RC, the reservoir effectively builds a complex manifold of trajectories by operating at the edge of chaos \cite{bertschingerRealTimeComputationEdge2004}. In our system, an analogous effect is driven by the nonlinear mixing of stretched random Fourier features. 

It is, however, crucial to define the boundaries of the phase-wrapping-induced improvements in expressivity. A sufficiently large $\alpha$ results in a regime of dense mode aliasing, in which the relative phases between terms in the interference sum are effectively randomised. In this limit, the output intensity becomes a high-dimensional inner product between pseudo-random vectors sampled from a wrapped phase manifold. The form of high-dimensional randomness inevitably results in a form of self-averaging - a well-known phenomenon in quantum extreme learning and "Random Kitchen Sinks", where it is termed \textit{exponential concentration} \cite{NIPS2008_0efe3284}. The consequence of this is that the overlap between any pair of output vectors saturates rapidly, and all readouts converge to their average. In other words, a complete loss of separability, but emerging as a consequence of statistical homogenization rather than periodicity.  

 While the full scope and applicability of phase wrapping to more complex tasks (such as time-series prediction) will be the subject of future work, the implications of the present findings are nevertheless both broad and immediate. Most prominently, they reinforce the critical role encoding plays in reservoir computing, and hint at a tantalizing possibility: dramatic gains in expressivity are possible at trivial costs once the role and structural complexity induced by the physical encoding scheme are fully taken into account and exploited. 

More broadly, exposing the Fourier representation implicit in phase encoding establishes immediate links with the broader domain of unconventional computing across multiple physical substrates. The quadratic mixing - and the phase interference that results - will be immediately recognizable to any quantum mechanics researcher. It reveals a near-exact alignment with \textit{Hamiltonian-encoded} quantum reservoir computers \cite{mccaulMinimalQuantumReservoirs2025} and quantum gate machine learning frameworks \cite{Xiong2025QELM,PhysRevA.103.032430}. From this perspective, the wrapping factor $\alpha$ effectively takes on the role of \textit{time evolution}. By such an analogy, domain extension becomes retrospectively obvious. These correspondences are likely to serve as a source of insight to \textit{both} quantum and classical RC, and further sharpen the question of quantum advantage in computing. Nascent efforts to identify and quantify this from the dynamical perspective \cite{mccaul_wave_2023} will surely benefit from the identification of novel analogs.  

Beyond this, it is - inevitably - impossible to ignore the geometric perspective. Any decoupling between an observable and an associated periodicity is \textit{definitionally} topological. Future investigations will focus on solidifying this conceptual bridge and mutually strengthening the capacities of each field, particularly due to potential connections with the rapidly expanding field of geometric deep learning \cite{bronsteinGeometricDeepLearning2021}.

Ultimately, our results further evince the unique advantages of encoding and processing data and information in the optical domain. In many ways, randomness, mixing, and Fourier encoding are the skeleton keys of unconventional computing. Their ease of access in the optical domain signals that the limits of their computational potential are yet to be reached. It is at times difficult to avoid the impression that the boundary of possibility is itself a periodic one, and that what goes around comes around.

\begin{acknowledgments}
This work was supported by the the UK Engineering and Physical Sciences Research Council (EPSRC, Grant No. EP/W028344/1), the European Union and the European Innovation Council through the Horizon Europe project QRC-4-ESP (Grant Agreement no. 101129663), and the UK Research and Innovation (UKRI) Horizon Europe guarantee scheme for the projects QRC-4-ESP (Grant No. 10108296). We acknowledge the use of the Lovelace HPC service at Loughborough University. GM and JSTG further express their gratitude to Barry Boldwater for his expert assistance in the development and sharpening of the theoretical analysis. We also thank Dr William Clements (ORCA Computing) for insightful discussions on the Fourier encoding in Quantum ML and Quantum Extreme Learning Machines settings. 
\end{acknowledgments}
\section*{Author Declarations}
G. Marcucci is Co-Founder and CEO of LumiAIres Ltd, a startup innovating in low-energy photonic AI. 
\section*{Authors Contributions}
JSTG and G. Marcucci designed the research and supervised the project. 
JSTG and G. McCaul developed the theoretical framework and interpretation. 
GT and JSTG developed the software and performed the main numerical calculations. 
All authors analyzed the data, interpreted the results, and contributed to writing and editing the paper. 
\section*{Code and Data Availability}
Numerical codes and data supporting this work are available from the corresponding authors following reasonable requests. 
\bibliography{refs}
\end{document}